\documentclass{article}

\usepackage{arxiv}

\usepackage[utf8]{inputenc} 
\usepackage[T1]{fontenc}
\usepackage{graphicx}
\usepackage{hyperref}
\usepackage{color}

\usepackage{mathtools}
\usepackage{bm}
\usepackage{amssymb}
\usepackage{amsmath} 
\usepackage{nicefrac}
\usepackage{todonotes}
\usepackage{braket}
\usepackage{siunitx}
\usepackage{booktabs}
\usepackage{multirow}
\usepackage{tikz}
\usepackage{tikz-network}
\usepackage{adjustbox} 
\usepackage{graphicx}
\usepackage{csquotes}
\usepackage{natbib}
\usepackage{wrapfig}
\usepackage[font=footnotesize]{caption, subcaption}
\usepackage{colortbl}
\colorlet{shadecolor}{gray!10}
\newcommand{\mycc}{\cellcolor{shadecolor}}

\DeclarePairedDelimiter\ceil{\lceil}{\rceil}
\DeclarePairedDelimiter\floor{\lfloor}{\rfloor}
\newcommand{\mat}[1]{\mathbf{#1}}
\newcommand{\vect}[1]{\bm{#1}}

\begin{document}

\title{A Comparative Study On Solving Optimization Problems With Exponentially Fewer Qubits}

\author{
    David Winderl, Nicola Franco, Jeanette Miriam Lorenz \\
    Fraunhofer Institute for Cognitive Systems IKS, \\
    Hansastrasse 32, 80686 \\
    Munich, Germany \\
  \texttt{\{name.middlename.surname\}@iks.fraunhofer.de}
}

\maketitle             

\begin{abstract}
Variational Quantum optimization algorithms, such as the Variational Quantum Eigensolver (VQE) or the Quantum Approximate Optimization Algorithm (QAOA), are among the most studied quantum algorithms. 
In our work, we evaluate and improve an algorithm based on VQE, which uses exponentially fewer qubits compared to the QAOA. 
We highlight the numerical instabilities generated by encoding the problem into the variational ansatz and propose a classical optimization procedure to find the ground-state of the ansatz in less iterations with a better or similar objective.
Furthermore, we compare classical optimizers for this variational ansatz on quadratic unconstrained binary optimization and graph partitioning problems.
\end{abstract}
\section{Introduction \& Related Work}\label{sec:introduction}

Since the emergence of Variational Quantum Algorithms (VQAs)~\citep{cerezo2021variational} and their subclasses, such as the Variational Quantum Eigensolver (VQE)~\citep{abrams1999quantum} and the Quantum Approximate Optimization Algorithm (QAOA)~\citep{farhi2014quantum}, the amount of research contributions towards the practical utility of quantum computers has increased every year. 
Applications of mathematical optimization are of particular interest, and QAOA is widely adopted in this context.
Although QAOA is an heuristic method, researchers are constantly advancing with slight improvements and precise adjustments to achieve better performance on narrow optimization problems.
 The most frequently studied problems are the Travelling Salesman Problem (TSP)~\citep{gavish1978travelling} and graph partitioning problems, such as MaxCut~\citep{commander2009maximum}. 
These NP-hard problems can be reduced to Quadratic Unconstrained Binary Optimization (QUBO)~\citep{glover2022quantum} or Ising~\citep{lucas2014ising} formulations.

As pointed out in~\cite{guerreschi2019qaoa}, because of the flaws of current Noisy Intermediate-Scale Quantum computers (NISQs), we may only see an actual speedup of large-scale optimization problems with QAOA for devices with several hundred qubits.
Hence recent contributions, have offered decomposition methods to overcome this issue and decompose the original problem to a size matching of NISQ devices.
In this context, decomposition methods, such as~\cite{li2021large} and~\cite{zhou2022qaoa}, offer solutions to overcome these limitations by splitting the original problem into smaller subproblems with sizes matching the requirements of NISQ hardware. 
In~\cite{li2021large}, after dividing a graph into two subgraphs that share common nodes, to obtain a possible candidate, the solutions of the respective subgraphs must overlap exactly. 
In this regard, the complexity of their approach increases with the number of common nodes, making it more difficult to find a good candidate solution.
Extending this, a new encoding strategy in~\cite{ranvcic2021exponentially} opens a new perspective as it requires exponentially fewer qubits to solve MaxCut with a VQE ansatz. 
To achieve this advantage, a continuous differentiable function maps the multi-dimensional binary optimization problem to a one-dimensional multi-modal continuous variable optimization problem.
In this work, after summarizing the approach of~\cite{ranvcic2021exponentially}, we highlight its limitations and suggest improvements in form of a proposal of new functions for encoding the problem. We conduct experiments on structured and random QUBO problems either in simulated and real quantum devices to demonstrate the viability of our improved approach.

\section{Background}\label{sec:study}

\subsection{Preliminaries}

Considering a weighted graph $\mathcal{G} = (V, E)$, we denote as $V = \{1, \dots, n\}$ the set of vertices, and $E = \{(i, j)\, |\, (i, j) \in V\times V\}$ the set of edges, with weights $w_{ij} \in \mathbb{R}$ for $(i, j) \in E$. 
In the weighted MaxCut problem, a cut is defined as a partition of the original set $V$ into two subsets. 
Given a binary vector $\vect{x} \in \mathbb{R}^{|V|}$, the cost function to be maximized is given by $\tilde{C}(\vect{x}) = \sum_{i, j} w_{ij} x_i (1-x_j)$. 
It represents the sum of weights of edges connecting points in the two different subsets, i.e. crossing the cut.
MaxCut is a specific case of a more general class of optimization problems that fall under the QUBO umbrella.
In the context of QUBO, one wants to minimize the following cost function: $f(\vect{x}) = \vect{x}^T\mat{Q}\vect{x}$, where $\vect{x} \in \{0, 1\}^n$ is a binary vector and $\mat{Q} \in \mathbb{R}^{n \times n}$ is a square matrix of real values.
We can rewrite a QUBO problem to its Ising variant by setting  $x_i = \frac{1}{2}v_i+\frac{1}{2}$ where $v_i\in\{-1, 1\}$ denotes an Ising spin, and further reduce it to MaxCut~\citep{Barahona1989}. 


\subsection{MaxCut with exponentially less qubits}

\begin{wrapfigure}{r}{0.4\textwidth}
    \vspace{-4em}
    \centering
    \begin{minipage}{0.4\textwidth}
    \centering\captionsetup[subfigure]{justification=centering}
    \includegraphics[width=\textwidth]{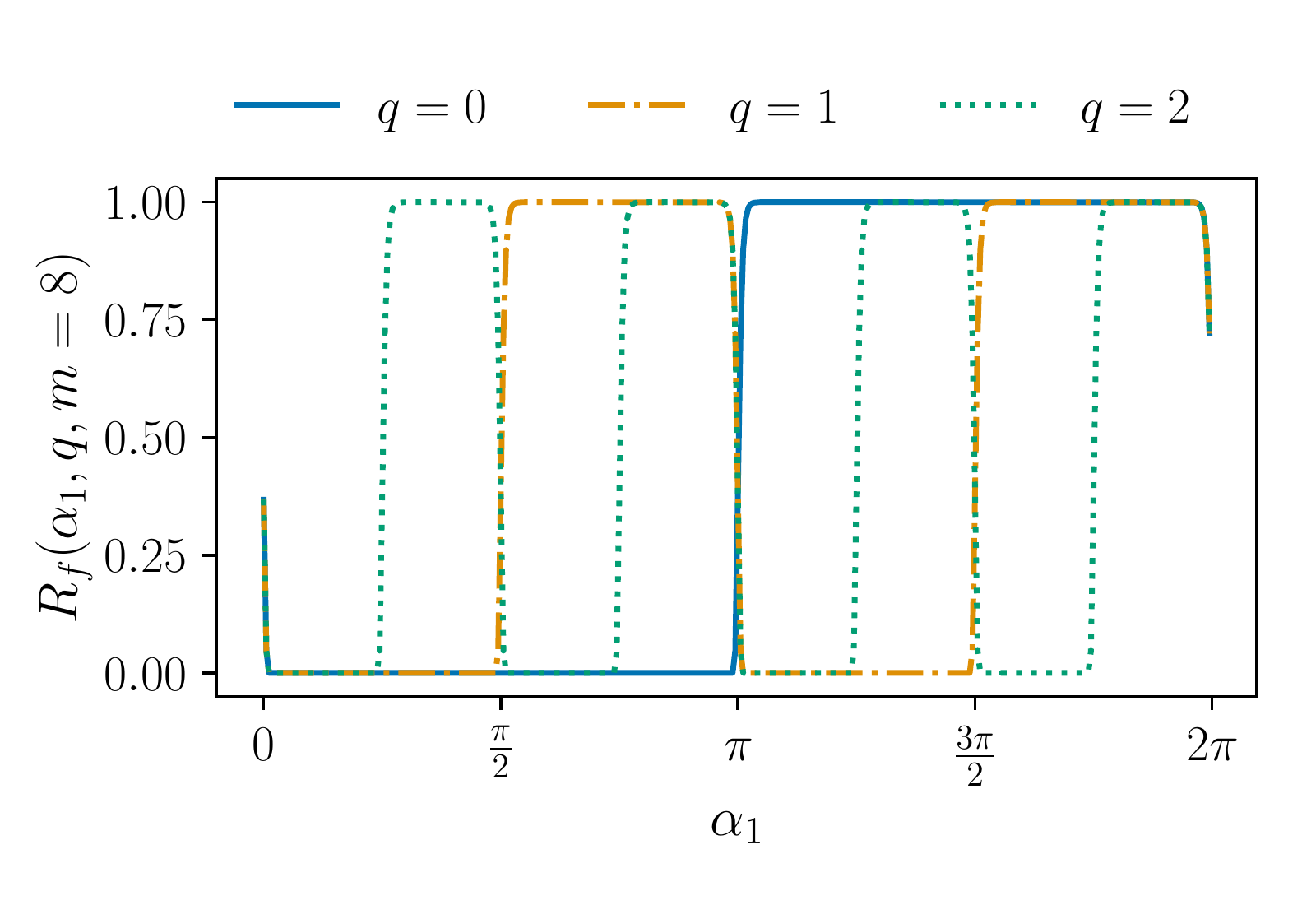}
    \vspace{-2.5em}
    \subcaption{$R_f(\alpha_1, q, m=6)$}
    \label{fig:rf_function}\par\vfill
    \vspace{-1em}
    \includegraphics[width=\linewidth]{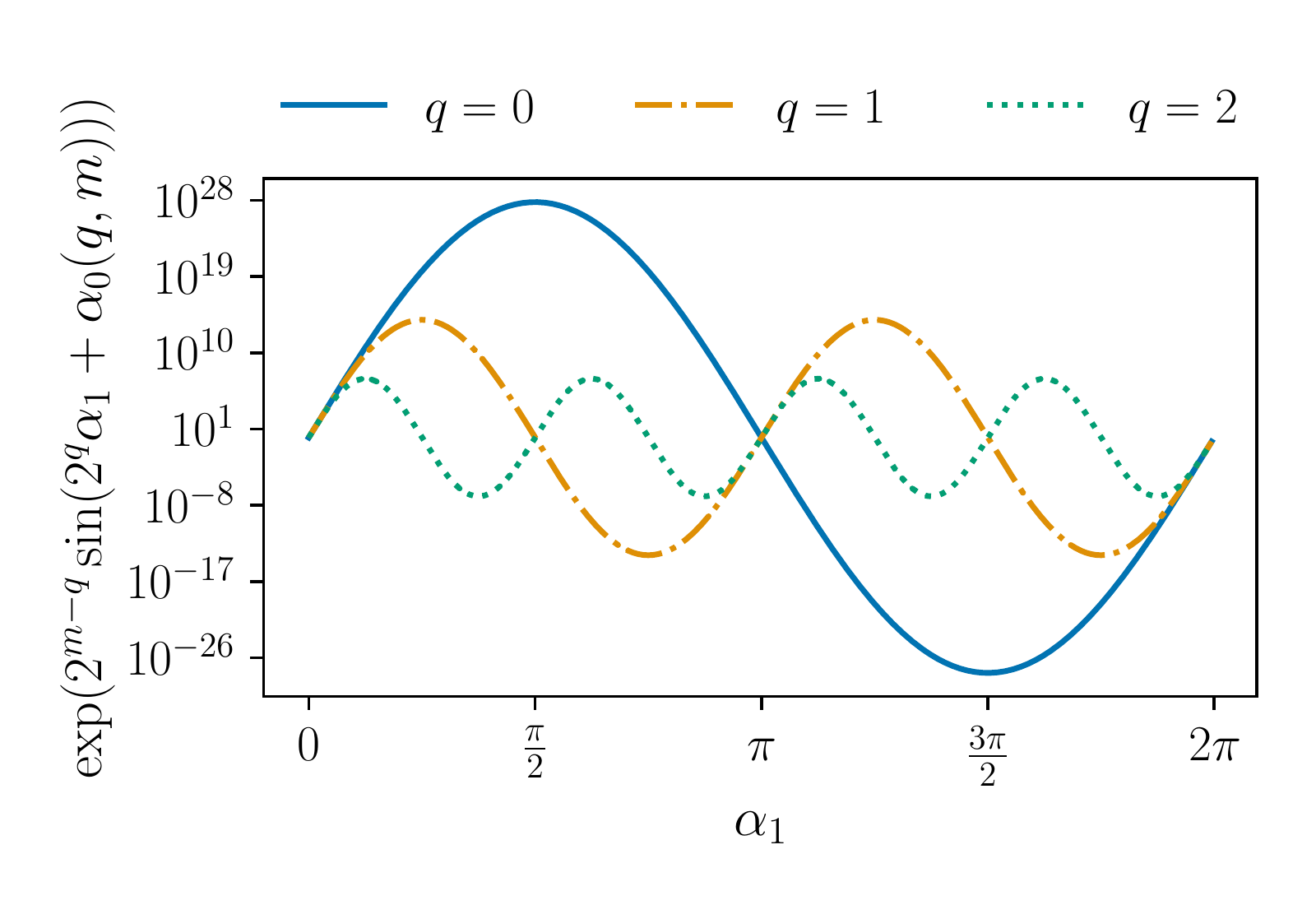}
    \vspace{-2.5em}
    \subcaption{exp$(2^{m-q}\sin (2^q \alpha_1 + \alpha_0(q, m)))$}
    \label{fig:rf_internal}
    \end{minipage}
    \caption{$R_f(\alpha_1, q, m=6)$ for $q\in\{0, 1, 2\}$.}
    \label{fig:rf}
\vspace{-1em}
\end{wrapfigure}

We start by considering that the number of cuts in MaxCut can be expressed as $N_{CUTS} = \frac{1}{4} \vect{v}^T L(\mathcal{G}) \vect{v}$,
where $\vect{v} \in \{-1, 1\}^{|V|}$ and $L(\mathcal{G})$ describes the Laplacian matrix of a graph. 
As presented in~\citet{ranvcic2021exponentially}, it is possible to map a continuous vector $\vect{\alpha} \in \mathbb{R}^k$, with $0 \leq \alpha_i \leq 2\pi \; \forall i\in\mathbb{N}^+$, and $1 \leq k \leq |V|-1$, into the spins vector $\vect{v}$ by utilizing a continuous differentiable function $R_f:\mathbb{R} \to [0, 1]$ defined as:
%
%
\begin{equation} \label{eq:r_f_default}
    R_f(\alpha_i, q, m) = \exp (-\exp(2^{m-q}\sin(2^q\alpha_i + \alpha_0(q, m)))),
\end{equation}
where $\alpha_0(q, m) = \arcsin (\nicefrac{\log_2(-\log_2(0.5))}{2^{m-q}})$ and $m, q\in\mathbb{N}^+_0$ with $m\geq |V|$ and $0 \leq q \leq |V|-2$.
In~\autoref{fig:rf_function} we see that for different values of $q$, the function alters between 0 and 1 with different frequencies when changing $\alpha_1$ from $0$ to $2\pi$. 
Following~\cite{ranvcic2021exponentially}, we fix $m$ and construct the vector of spins $\vect{v}$ as output of $R_f$ by enumerating all $q$ values: $\vect{v}(\alpha_i, m) = (e^{i\pi R_f(\alpha_i, 0, m)}, \dots e^{i\pi R_f(\alpha_i, d_i, m)})$.
One can create such vectors $\vect{v}\in \{0, 1\}^{d_i}$ for each $\alpha_i$, where $\sum_{i =1}^{k}d_i = |V|-1$, which leads towards the diagonal gate $U(\vect{\alpha})$, specified as follows:
\begin{align*}
    U(\vect{\alpha}) &= \text{diag}(v(\alpha_1), \dots v(\alpha_{k}), 1, \dots 1) \\
    &= \text{diag}(e^{i\pi R_f(\alpha_1, 0, m)}, \dots , e^{i\pi R_f(\alpha_{k}, d_k, m)}, 1, \dots 1),
\end{align*}
where the size of $U(\vect{\alpha})$ is $2^{\ceil{\log_2(|V|)}}$. 
$U(\vect{\alpha})$ can be realized using multiplexor gates as outlined in \cite{Shende2004}.
Next, assuming that the Laplacian is a $d$-sparse matrix and can be effectively embedded on a quantum device~\cite{pothen1990partitioning}\footnote{Note that while this is possible, we used a less efficient method and just decomposed the matrix as a tensor-product of pauli matrices}, the variational ansatz is given by:
\begin{equation}\label{eq:ansatz}
    N_{CUTS} = \frac{1}{4} \vect{v}^TL(\mathcal{G})\vect{v} = 2^{n-2}\bra{0}HU(\vect{\alpha})\mathcal{L}(\mathcal{G})U(\vect{\alpha})H\ket{0},
\end{equation}
where, $H$ describes the Hadamard transform and $L(\mathcal{G})$ is represented as a sum of tensor products of unitary matrices, and denoted as $\mathcal{L}(\mathcal{G})$ in such form.

The $R_f$ function in~\autoref{eq:r_f_default} serves as a key point for the exponential qubits reduction presented in~\citet{ranvcic2021exponentially}. 
Despite this advantage, the function is highly numerically unstable for larger graphs. 
To make this clear, let us assume a graph with $|V| = 6$, then set $m=6$ (since according to~\citet{ranvcic2021exponentially}, $m \geq |V|$) and compute $R_f$ for $\alpha_1= \frac{\pi}{2}$ and $q=0$.
In~\autoref{fig:rf_internal} one can see, that partially evaluating $\exp (2^{6-0} \cdot \sin{(2^0\frac{\pi}{2} + \alpha_0(0, 6))}) \approx \num{6.2e+27}$ already results in a large number. 
Hence computing this function for large graphs with not enough dimensions for $\vect{\alpha}$ is not feasible.
\section{Method}\label{sec:method}

\subsection{Improvements on $R_f$}\label{subsec:improvements_on_rf}

As previously discussed, $R_f$ creates alternating plateaus on either zero or one which are then used to enumerate all possible binary combinations of length $2^{|V|}$.
To achieve the same behavior of $R_f$, with a more numerically stable function, we consider a modified version of the sawtooth wave.

\begin{wrapfigure}{l}{0.4\textwidth}
\vspace{-2em}
    \centering
    \includegraphics[width=0.4\textwidth]{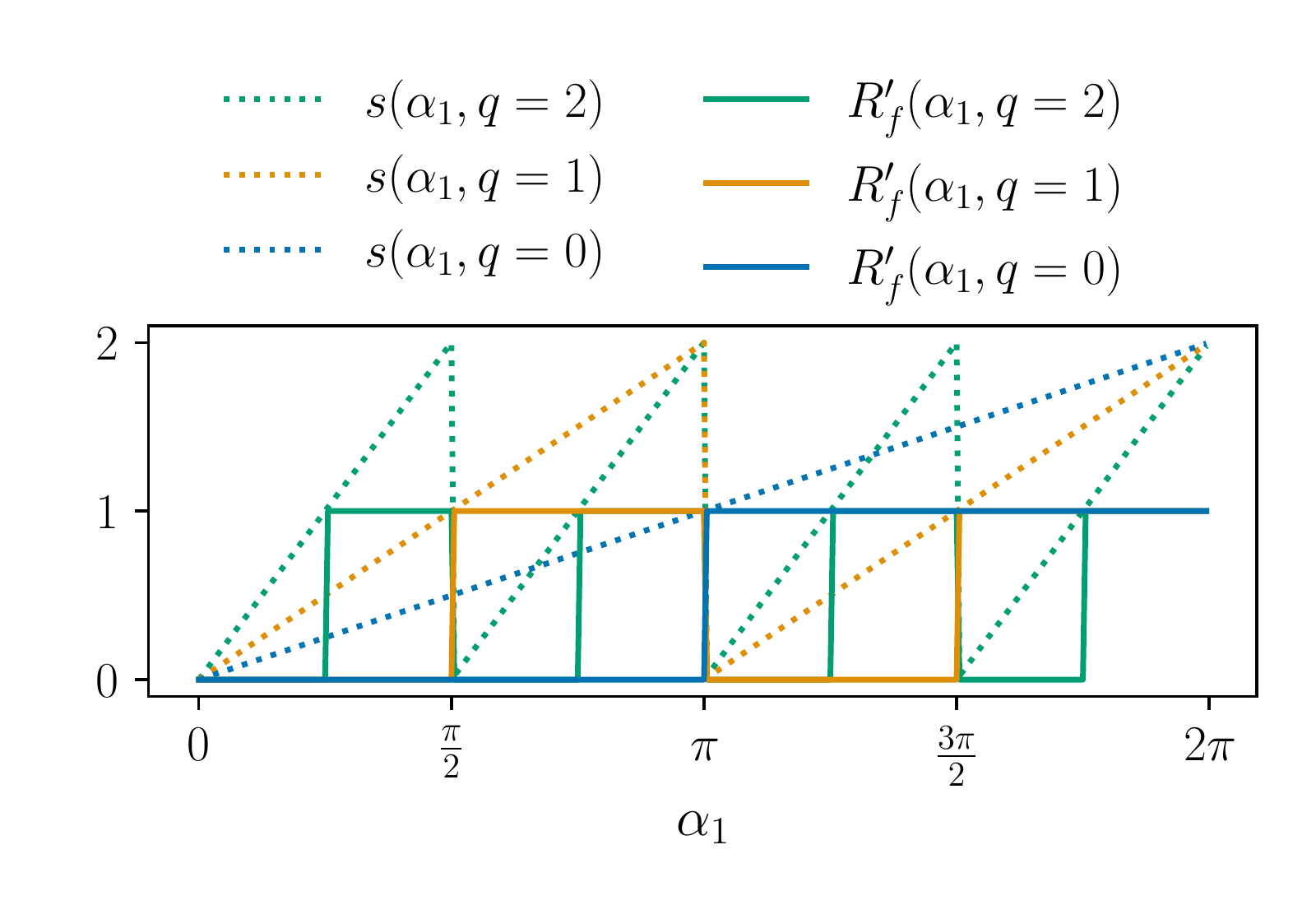}
    \vspace{-2em}
    \caption{$s(\alpha_1, q)$ and $R^\prime_f(\alpha_1, q)$ for $q\in \{0, 1, 2\}$}
    \label{fig:rf_prime}
\vspace{-2em}
\end{wrapfigure}
For this reason, we introduce our proposed version: 
\begin{equation}
    s(\alpha_i, q) = 1-\frac{2}{\pi}\arctan{(\cot ({2^{q-1}\alpha_i}))}.
\end{equation}
Hence, we can mimic $R_f$ with $\floor{s(\alpha_i, q)}$, where $\floor{\cdot}$ represents the floor function.
Since the floor function is nonlinear, we can approximate it by precalculating $\bar{s}(\alpha_i, q)$ for a given $\alpha_i$ and $q$, and, in the next step, calculating:
\begin{equation}\label{eq:improved_rf}
    R^\prime_f(\alpha_i, q) = \bar{s}(\alpha_i, q) - \frac{1}{2}s(2^{1-q}\pi\bar{s}(\alpha_i, q)).
\end{equation}
In~\autoref{fig:rf_prime}, we show both $s(\alpha_i, q)$ and $R^\prime_f(\alpha_i, q)$ for a given $\alpha_1$.
In the end, we observe two advantages of the proposed function: (i) the values of $s(\alpha_i, q)$ are contained within the interval $[0, 2]$, which makes $R^\prime_f(\alpha_i, q)$ numerically stable, and (ii) the reduction in the number of hyperparameters due to the removal of $m$.

\subsection{Alternating Optimization Procedure (AltOpt)}\label{subsec:altopt}

Here, we present our algorithm based on the alternating optimization strategy \citep{Bezdek2002} to find the ground state for the variational ansatz in~\autoref{eq:ansatz}. 
Essentially, this procedure uses the fact that the maximum number of variables in $\vect{\alpha}$ can be determined as $|V| - 1$. 
Given that, one can decide per diagonal entry of $U(\vect{\alpha})$ to set it to $-1$ or $1$. In the space of our continuous variable $\alpha_i$, this corresponds to the values $\frac{\pi}{2}$ or $\frac{3\pi}{2}$, respectively.
Thus we propose the following procedure to minimize ~\autoref{eq:ansatz}:
\begin{enumerate}
    \item Initialize all diagonal entries to 1 (set their corresponding variable to $\frac{\pi}{2}$).
    \item Travel the diagonal from the top left to bottom right, selecting the value with lower energy each time. 
    \item Restart from the top left until the objective no longer improves.
\end{enumerate}
The proposed optimization method has several good properties: hyperparameter-free, faster convergence, and less dependence on the initial choice of the parameters. Nevertheless, it is specific for the variational ansatz of \autoref{eq:ansatz}.
\section{Experiments}\label{sec:experiments}

In this section, we compare the algorithm of~\citet{ranvcic2021exponentially} on three optimization problems: randomly generated QUBOs, TSP and MaxCut instances.
To this end, we evaluate the performance with respect to a selection of classical optimizers: Nakanishi-Fujii-Todo algorithm (\textbf{NFT})~\citep{Nakanishi_2020}\footnote{with a maximum of 200 iterations and 500 function evaluations.}, Basinhopping (\textbf{BH})~\citep{Olson2012}\footnote{with initial state $\pi$, stepsize of $2\pi$, an interval to update the stepsize of 10, a maximum of 200 iterations and a local minimizer COBYLA with a rhobeg of $\nicefrac{\# Variables}{2^{|V|-1}}$.}, as well as the Genetic Algorithm (\textbf{GA}) with the settings described in~\cite{ranvcic2021exponentially}, and our proposed new method \textbf{AltOpt}.
For a smaller subset of random QUBOs we conduct the study on real hardware. 
Lastly, we compare our energy to the \textbf{Tabu} Solver~\footnote{dwave-tabu \url{https://github.com/dwavesystems/dwave-tabu}} of \textit{Dwave}~\cite{Palubeckis2004}, which is also a search heuristic to solve QUBO or Ising Problems of a larger size.
The number of iterations is defined as a repetition loop between the classical and the quantum solver.

\clearpage
\subsection{Random QUBOs on simulated hardware}

\begin{wraptable}{r}{0.5\textwidth}
\vspace{-1em}
\centering
    \caption{Evaluation on random QUBO problems of different sizes. \label{tab:random_qubos_classic}}
    \vspace{-0.5em}
    \begin{adjustbox}{width=0.5\textwidth,center}
        \begin{tabular}{l|rrrrr|rrrr}
        \toprule
            \multirow{2}{*}{\textbf{Size}} &\multicolumn{5}{c}{\textbf{Energy} $\downarrow$} &\multicolumn{4}{c}{\textbf{\# Iterations $\downarrow$}} \\
                          &Tabu &GA &AltOpt & BH &NFT           & GA & AltOpt & BH &NFT    \\
            \midrule
            \textbf{15}   &\mycc-8,6   &\textbf{-8,4}   & -8,2    & -5,6 & -4,5      & 228  &\textbf{43} & 195 & 314 \\
            \textbf{31}   &\mycc-25,4  & -22,8  &\textbf{-24,5}  & -11,7 & -5,9      & 345  &\textbf{119} & 201 & 394 \\
            \textbf{63}   &\mycc-72,9  & -9,3   &\textbf{-71,1}  & -14,4 & -43,7     & 399  & 298 & \textbf{201} & 402 \\
            \textbf{127}  &\mycc-203,9 & 1,1    &\textbf{-199,4} & -25,6 & -46,0     & 401  & 806 & \textbf{201} & 408 \\
        \bottomrule
        \end{tabular}
    \end{adjustbox}
    \vspace{-1em}
\end{wraptable}

We generated random QUBO matrices of different size to test the average performance of each algorithm.
For each size, we average over densities (how many non zero elements) in $[0.1, 0.2, \dots 0.9]$ with 20 samples per size and density. 
All elements of matrix $Q$ are sampled from a normal distribution with mean 0 and variance 1. 
Then, we evaluate the symmetric matrix $\frac{1}{2}(Q + Q^T)$.
The results are presented in~\autoref{tab:random_qubos_classic}.
We find that for random QUBOs, AltOpt shows comparable results to the Tabu-Solver in terms of energy. 
GA demonstrates similar energy to AltOpt for small instances, nevertheless for instances of 63 and 127 the energy is much larger. 
Both Basinhopping and NFT have worse results on average.

\subsection{Travelling Salesman Problem}

\begin{table}[htb!]
\vspace{-1em}
    \centering
    \caption{Evaluation on randomly generated TSP instances, for different sizes. We are taking the average over 20 samples per size.}\label{tab:tsp_solutions}
    \begin{adjustbox}{width=\textwidth,center}
    \begin{tabular}{l|rrrrr|rrrrr|rrrr}
    \toprule
    \multirow{2}{*}{\textbf{Size}} & \multicolumn{5}{c}{\textbf{Energy} $\downarrow$} &\multicolumn{5}{c}{\textbf{Feasibility} $\uparrow$} &\multicolumn{4}{c}{\textbf{\#  Iterations} $\downarrow$} \\
                        &Tabu &GA &AltOpt &BH &NFT &Tabu &GA &AltOpt &BH &NFT &GA &AltOpt &BH &NFT    \\
                                   \midrule
    \textbf{3}               &\mycc-439,9    &\textbf{-439,9}     &\textbf{-439,9}     &-410,6    &-346,9      &\mycc1 &\textbf{1}      &\textbf{1}     &0.65 &0,45   &199    &\textbf{18}     &240    &344    \\
    \textbf{5}               &\mycc -771,2   &-721,5     &\textbf{-727,9}     &280,3     &902,7       &\mycc1 &0.75   &\textbf{0.95}  &0    &0      &339    &\textbf{74}     &201    &406    \\
    \textbf{7}               &\mycc -1158,9  &-666,35    &\textbf{-1036,5}    &1857,3    &406,8       &\mycc1 &0      &\textbf{1}     &0    &0      &404    &\textbf{141}    &201    &400    \\
    \textbf{9}               &\mycc -1507,4  &1650,4     &\textbf{-1352,6}    &27171,4   &10795,1     &\mycc1 &0      &\textbf{0.95}  &0    &0      &405    &241    &\textbf{201}    &408    \\
    \textbf{11}              &\mycc -1898,9  &10503,5    &\textbf{-1627,8}    &56191,7   &20480,5     &\mycc1 &0      &\textbf{1}     &0    &0      &405    &359    &\textbf{201}    &408    \\
    \bottomrule
    \end{tabular}
    \end{adjustbox}
\end{table}

Here, we test the variational ansatz on problems, that provide a inherent structure. 
For this reason, we took the TSP, which comes along with a QUBO or Ising formulation already~\cite{lucas2014ising, Glover2018}.
For a fair comparison, we evaluate the results in terms of energy, number of iterations and normalized feasibility. 
A solution is feasible if it produces a Hamiltonian path through the graph. 
Given the results in~\autoref{tab:tsp_solutions}, we see a similar pattern of random QUBOs. 
Furthermore, it is worth noting that AltOpt returns a substantial percentage of feasible solutions. 
On the one side, GA, BH, and NFT struggle to explore the solution landscape and stop at unfeasible local minima.
On the other side, the maximum number of iterations set at 400 for GA and NFT and 200 for BH limits their search capabilities.
\subsection{Max-Cut}

\begin{wraptable}{l}{0.5\textwidth}
\vspace{-1em}
    \caption{Evaluation of MaxCut on $d$-regular graphs for different sizes.}
    \vspace{-0.5em}
    \label{tab:maxcut_exvaluation}
    \centering
    \begin{adjustbox}{width=0.5\textwidth,center}
    \begin{tabular}{l|rrrr|rrrr}
        \toprule
        \multirow{2}{*}{\textbf{Size}} & \multicolumn{4}{c}{\textbf{Cut Value} $\uparrow$}     & \multicolumn{4}{c}{\textbf{\# Iterations} $\downarrow$}      \\
                                       & GA & AltOpt  & BH & NFT     & GA & AltOpt  & BH & NFT    \\\midrule
        \textbf{16}                 & 32,2   & \textbf{32,3}   & 30,8        & 31,5    & 219,4  & \textbf{45,2}   & 193       & 293 \\
        \textbf{32}                 & 125,1  & 127,4   & \textbf{127,8}       & 127,0  & 340,7  & \textbf{120,7}  & 201       & 394 \\
        \textbf{64}                 & 505,0  & \textbf{512,7}  & 500,5       & 511,4  & 404,4  & 355,8  & \textbf{201}          & 402 \\
        \textbf{128}                & 2033,2 & \textbf{2050,2} & 2041,7      & 2040,7 & 404,9  & 1016,6 & \textbf{201}          & 408 \\
        \bottomrule
    \end{tabular}
    \end{adjustbox}
    \vspace{-1em}
\end{wraptable}

Here, we conduct test on different instances of MaxCut on $d$-regular graphs.
For the degree of the graph $d$, we used 9 uniformly distributed values between 0 and the size of the graph.
We use 20 graphs per factor of $d$, assign uniformly random edge weights in the interval $[0,5]$ and averaged the resulting Cut Value.
The results are shown in~\autoref{tab:maxcut_exvaluation}. 
Interestingly, all algorithms tend to perform similarly.
We account this and the difficulties to solve QUBO towards the fact, that the reduction of QUBO-Matrices to MaxCut does not provide $d$-regular graphs.

\subsection{Random QUBOs on real hardware}

\begin{wraptable}{r}{0.5\textwidth}
\vspace{-1em}
    \caption{Evaluation on random QUBO problems with quantum hardware. We run 3 instances per algorithm and average the energy and number of iterations.}
    \vspace{-0.5em}
    \label{tab:quantum_hardware}
    \centering
    \begin{adjustbox}{width=0.5\textwidth,center}
    \begin{tabular}{ll|rrrr|rrr}
    \toprule
    \multirow{2}{*}{\textbf{Size}} &\multirow{2}{*}{\textbf{Density}} &\multicolumn{4}{c}{\textbf{Energy} $\downarrow$} &\multicolumn{3}{c}{\textbf{\# Iterations} $\downarrow$} \\
    & &Tabu &AltOpt &BH &NFT &AltOpt &BH &NFT \\
    \midrule
    \multirow{3}{*}{\textbf{31}} 
        &0,1 &\mycc-10,8 &-1,1 &\textbf{-3,8} &0,4 &\textbf{83} &201 &408 \\
        &0,5 &\mycc-24,8 &\textbf{-4,3} &-0,3 &14,5 &\textbf{63} &201 &396 \\
        &0,9 &\mycc-27,8 &-1.3 &\textbf{-1,7} &12,9 &\textbf{70} &201 &395 \\
    \bottomrule
    \end{tabular}
    \end{adjustbox}
\vspace{-1em}
\end{wraptable}

Finally, we conduct experiments on gate-based quantum computers. 
In this context, we access the 7 qubits systems: lagos and jakarta, with Qiskit~\citep{Qiskit} from IBM Quantum\footnote{\copyright\ IBM Quantum https://quantum-computing.ibm.com/}, both based on the Falcon r5.11 architecture, and having a quantum volume of 32 and 16, respectively.
In \autoref{tab:quantum_hardware}, we present results for a random QUBO problem of size 31 over 3 density levels. 
The performance of AltOpt, BH and NFT is lower in all three tests, with AltOpt and BH performing better in general.
This behaviour has to be attribute at the intrinsic noise generated by the system, where to improve the results quality, a larger number of shoots will be required, as demonstrated in \cite{ranvcic2021exponentially}, where 8192 was used instead of 1024 of our evaluation.
\section{Conclusion}

In this work, we evaluate \cite{ranvcic2021exponentially} on a variety of QUBO-Problems. Additionally, we provide an improved version $R^\prime_f$ of the previously proposed function $R_f$. It has been shown, that $R^\prime_f$ is numerically stable and robust against larger inputs. 
For random QUBOs our optimization method is comparable to the Tabu optimizer, and outperforms state-of-the-art variational methods. 
Nevertheless, we show similar results on MaxCut for $d$-regular graphs.

\subsubsection*{Acknowledgment.}
The project/research is supported by the Bavarian Ministry of Economic Affairs, Regional Development and Energy with funds from the Hightech Agenda Bayern.

\bibliographystyle{unsrtnat}
\bibliography{bibliography}

\end{document}